\documentclass[11pt]{article}
\usepackage[utf8]{inputenc}
\usepackage{amsfonts}
\usepackage{amsmath}
\usepackage{amssymb}
\usepackage{indentfirst}
\usepackage{graphicx}
\usepackage[colorlinks]{hyperref}
\usepackage{cite}
\usepackage{colortbl}

\usepackage{microtype}
\usepackage[normalem]{ulem}

\numberwithin{equation}{section}
\allowdisplaybreaks

\begin{document}

\begin{titlepage}
\vspace{3cm}

\baselineskip=24pt

\begin{center}
\textbf{\LARGE{Three-dimensional non-relativistic supergravity and torsion}}
\par\end{center}{\LARGE \par}

\begin{center}
	\vspace{1cm}
	\textbf{Patrick Concha}$^{\ast}$,
	\textbf{Lucrezia Ravera}$^{\dag, \ddag}$,
	\textbf{Evelyn Rodríguez}$^{\star}$,
	\small
	\\[5mm]
    $^{\ast}$\textit{Departamento de Matemática y Física Aplicadas, }\\
	\textit{ Universidad Católica de la Santísima Concepción, }\\
\textit{ Alonso de Ribera 2850, Concepción, Chile.}
	\\[2mm]
	$^{\dag}$\textit{DISAT, Politecnico di Torino, }\\
	\textit{ Corso Duca degli Abruzzi 24, 10129 Torino, Italy.}
	\\[2mm]
	$^{\ddag}$\textit{INFN, Sezione di Torino, }\\
	\textit{ Via P. Giuria 1, 10125 Torino, Italy.}
	\\[2mm]
	$^{\star}$\textit{Instituto de Matemática (INSTMAT), Universidad de Talca, }\\
	\textit{Casilla 747, Talca 3460000, Chile.}
	\\[5mm]
	\footnotesize
	\texttt{patrick.concha@ucsc.cl},
	\texttt{lucrezia.ravera@polito.it},
	\texttt{evelyn.rodriguez@utalca.cl},
	\par\end{center}
\vskip 26pt
\begin{abstract}

In this paper we present a torsional non-relativistic Chern-Simons teleparallel (super)gravity theory in three spacetime dimensions. We start by developing the non-relativistic limit of the purely bosonic relativistic teleparallel Chern-Simons formulation of gravity. On-shell the latter yields a non-Riemannian setup with non-vanishing torsion, which, at non-relativistic level, translates into a non-vanishing spatial torsion sourced by the cosmological constant. Then we consider the three-dimensional relativistic $\mathcal{N}=2$ Chern-Simons supergravity theory and obtain its non-relativistic counterpart by exploiting a Lie algebra expansion method. The non-relativistic supergravity theory is characterized, on-shell, by a non-vanishing spatial super-torsion, again sourced by the cosmological constant.

\end{abstract}
\end{titlepage}\newpage {} {\baselineskip=12pt }

\section{Introduction}\label{sec1}

The introduction of torsion in a gravity theory can be performed through the teleparallel formulation of gravity \cite{Hayashi:1979qx,Kawai:1993kvr,deAndrade:1997gka,Sousa:2000bc,DeAndrade:2000sf}, which is considered to be equivalent to General Relativity. Nevertheless, teleparallel gravity is characterized by a vanishing curvature and a non-vanishing torsion describing a non-Riemannian geometry denoted as Weizenböck geometry. In three spacetime dimensions, the role of torsion in black hole solutions and boundary dynamics has been of particular interest \cite{Garcia:2003nm,Mielke:2003xx,Blagojevic:2003vn,Blagojevic:2003uc,Blagojevic:2003wn,Blagojevic:2013bu,Adami:2020xkm}. As it was shown in \cite{Garcia:2003nm,Mielke:2003xx,Blagojevic:2003vn}, torsional three-dimensional gravity theory possesses a BTZ-like black hole solution whose thermodynamic properties have been studied in \cite{Blagojevic:2006jk,Blagojevic:2006hh,Blagojevic:2006nf}. Moreover, both teleparallel gravity and AdS gravity have identical asymptotic symmetries given by two copies of the Virasoro algebra \cite{Blagojevic:2003uc}. Interestingly, three-dimensional teleparallel gravity can be formulated as a gauge theory using the Chern-Simons (CS) formalism and considering the so-called teleparallel algebra \cite{Caroca:2021njq}. The teleparallel CS gravity action can be alternatively recovered as a particular case of the Mielke-Baelker gravity model \cite{Mielke:1991nn} by fixing the Mielke-Baelker parameters. Remarkably, the CS approach allows us to define a supersymmetric teleparallel gravity in the presence of $\mathcal{N}=p+q$ supercharges which in the vanishing cosmological constant limit reproduces the $\mathcal{N}$-extended Poincaré supergravity theory \cite{Caroca:2021njq}.

On the other hand, non-relativistic (NR) theories have received a growing interest due to their relation to condensed matter systems \cite{Son:2008ye,Balasubramanian:2008dm,Kachru:2008yh,Bagchi:2009my,Bagchi:2009pe,Christensen:2013lma,Christensen:2013rfa,Hartong:2014oma,Hartong:2014pma,Hartong:2015wxa,Taylor:2015glc} and NR effective field theories \cite{Hoyos:2011ez,Son:2013rqa,Abanov:2014ula,Geracie:2015dea,Gromov:2015fda}. A NR theory can be obtained by taking the limit $c\rightarrow \infty$, where $c$ is the speed of light, of a relativistic theory. At the gravity level, the corresponding NR counterpart of the AdS spacetime is the Newton-Hooke symmetry which in the vanishing cosmological constant limit reproduces the Galilei symmetry \cite{Bacry:1968zf, Bacry:1986pm,Aldrovandi:1998im,Gao:2001sr,Gibbons:2003rv,Brugues:2006yd,Alvarez:2007fw}. In three spacetime dimensions, the CS formalism allows us to construct NR gravity actions whose underlying symmetry can be obtained as a NR limit of a relativistic algebra. However, it is necessary to consider additional $\mathfrak{u}\left(1\right)$ generators in order to avoid infinities and degeneracy in the NR limit. Indeed, the Poincaré algebra requires the presence of two extra $\mathfrak{u}\left(1\right)$ generators in order to apply an appropriate and well-defined NR limit with a non-degenerate bilinear form and reproduces the extended Bargmann algebra \cite{Grigore:1993fz,Bose:1994sj,Duval:2000xr,Jackiw:2000tz,Papageorgiou:2009zc}. In the presence of a cosmological constant, the NR version of the AdS$\oplus\mathfrak{u}\left(1\right)^{2}$ symmetry corresponds to an extension of the Newton-Hooke symmetry \cite{Papageorgiou:2010ud,Duval:2011mi,Hartong:2016yrf,Duval:2016tzi}.

The formulation of a NR supergravity theory has only been approached recently and remains as a challenging task mainly motivated by the diverse applications of these models in the context of holography and relativistic field theory \cite{Andringa:2013mma,Bergshoeff:2015ija,Bergshoeff:2016lwr,Ozdemir:2019orp,Concha:2019mxx,Concha:2020eam}. One way to circumvent the difficulty to establish a well-defined NR limit in the presence of supersymmetry is through the Lie algebra expansion methods of \cite{Hatsuda:2001pp,deAzcarraga:2002xi,Izaurieta:2006zz,deAzcarraga:2007et}, which allows us to obtain the respective NR version of a Lie (super)algebra \cite{deAzcarraga:2019mdn,Ozdemir:2019tby,Gomis:2019nih,Concha:2020tqx,Concha:2021jos}. In particular, the semigroup expansion (S-expansion) method, introduced in \cite{Izaurieta:2006zz} and further developed in \cite{Caroca:2011qs,Andrianopoli:2013ooa,Artebani:2016gwh,Ipinza:2016bfc,Penafiel:2016ufo,Inostroza:2018gzd}, allows us not only to obtain the corresponding NR (super)algebra but also provides us with the non-vanishing components of the invariant tensor of the expanded (super)algebra. The S-expansion procedure consists
in combining the elements of a semigroup $S$ with the structure constant of a Lie (super)algebra
$\mathfrak{g}$. The new Lie (super)algebra $\mathfrak{G} = S \times \mathfrak{g}$ is said to be an S-expanded (super)algebra.\footnote{For a concise review of this expansion method we refer the reader to, e.g., Appendix A of Ref. \cite{Concha:2020tqx}.} Further developments of the Lie algebra expansion method in the NR context has also been explored in \cite{Bergshoeff:2019ctr,Concha:2019lhn,Penafiel:2019czp,Romano:2019ulw,Bergshoeff:2020fiz,Kasikci:2020qsj,Concha:2020sjt,Fontanella:2020eje,Concha:2020ebl,Kasikci:2021atn}.

In this work, we explore the NR counterpart of the teleparellel CS (super)gravity introduced in \cite{Caroca:2021njq}. In particular, we are interested in studying the role of a non-vanishing torsion in a non-relativistic environment. To this end, we first explore the NR limit of a particular $U\left(1\right)$-enlargement of the teleparallel algebra introduced in \cite{Caroca:2021njq}. We obtain a torsional NR algebra by applying a proper Inönü-Wigner contraction to the [teleparallel]$\oplus\mathfrak{u}\left(1\right)^{2}$ algebra. The corresponding NR gravity theory contains a non-vanishing NR torsion in which the cosmological constant can be seen as a source for the curvature of the spatial component of the vielbein. The supersymmetric extension of our theory is also presented by applying the S-expansion to the $\mathcal{N}=2$ teleparallel superalgebra presented in \cite{Caroca:2021njq}. As it was shown in \cite{Gomis:2019nih,Concha:2020tqx}, a particular semigroup allows to reproduce a Galilean expansion providing the respective NR (super)algebra and (super)gravity CS action. Interestingly, the novel torsional NR (super)gravity theory and the extended Bargmann (super)gravity theory \cite{Bergshoeff:2016lwr} are related through a vanishing cosmological constant limit as their respective relativistic versions. It is important to emphasize that along this work, unlike the Newton-Cartan (super)gravity theory with torsion \cite{Bergshoeff:2015ija,Bergshoeff:2017dqq,VandenBleeken:2017rij}, we shall refer to a torsional NR (super)gravity theory when the spatial component of the NR (super) torsion is non-zero. Thus, a vanishing torsion in our context is not intended as a zero torsion condition as in the torsional Newton-Cartan geometry \cite{Bergshoeff:2014uea,Festuccia:2016awg,Bergshoeff:2017pqf}.

The paper is organized as follows: In Section \ref{sec2} we briefly review the teleparallel CS gravity theory defined in three spacetime dimensions. Section \ref{sec3} and \ref{sec4} contain our main results. In Section \ref{sec3} we first present the torsional NR gravity theory by applying a NR limit to the teleparallel CS gravity. Section \ref{sec4} is devoted to the construction of a supersymmetric extension of the torsional NR gravity theory considering the S-expansion procedure. Section \ref{sec5} concludes our work with some discussions about future developments.

\section{Three-dimensional teleparallel formulation of gravity}\label{sec2}

At the relativistic level, the inclusion of a non-vanishing torsion in a three-dimensional gravity theory can be done through the teleparallel formulation of gravity \cite{Kawai:1993kvr,deAndrade:1997gka,DeAndrade:2000sf,Blagojevic:2003uc,Blagojevic:2003vn}. Interestingly, teleparallel gravity can be formulated as a CS action invariant under a particular Lie algebra which has been denoted as teleparallel algebra \cite{Caroca:2021njq}. Such algebra, which is isomorphic to two copies of the $\mathfrak{so}\left(2,1\right)$ algebra \cite{Parsa:2018kys}, is spanned by the set of generators $\left(J_A,P_A\right)$ and satisfies 
\begin{eqnarray}
\left[J_A,J_B\right]&=&\epsilon_{ABC} J^{C} \,, \notag \\
\left[J_A,P_B\right]&=&\epsilon_{ABC} P^{C} \,, \notag \\
\left[P_A,P_B\right]&=&-\frac{2}{\ell}\epsilon_{ABC} P^{C} \,, \label{telea} 
\end{eqnarray}
where $A,B,C=0,1,2$ are Lorentz indices and $\epsilon_{ABC}$ is the Levi-Cevita tensor defined in three spacetime dimensions. Here, the $\ell$ parameter is related to the cosmological constant $\Lambda$ through $\Lambda \propto - \frac{1}{\ell^2}$. In the vanishing cosmological constant limit $\ell\rightarrow \infty$, analogously to the $\mathfrak{so}\left(2,2\right)$ symmetry, we obtain the Poincaré algebra. Let us note that the Lie algebra \eqref{telea} admits the following non-vanishing components of the invariant tensor:
\begin{equation}
    \langle J_{A} J_{B} \rangle = \alpha_0 \,\eta_{A B}\,,\qquad \langle J_{A} P_{B} \rangle = \alpha_1 \,\eta_{A B}\,,\qquad \langle P_{A} P_{B} \rangle = -\frac{2\alpha_1}{\ell} \,\eta_{A B}\,,\label{IT1}
\end{equation}
where $\alpha_0$ and $\alpha_1$ are arbitrary constants and $\eta_{AB}$ is the flat Minskowski metric. On the other hand, the gauge connection one-form $A$ reads
\begin{equation}
    A= W^{A}J_{A}+E^{A} P_{A}\,,\label{A1f}
\end{equation}
where $W^{A}$ is the spin connection and $E^{A}$ denotes the vielbein satisfying $E^{A}_{\mu}E^{B}_{\nu}g^{\mu\nu}=\eta^{AB}$. The curvature two-form $F=dA+\frac{1}{2}[A,A]$ reads
\begin{equation}
    F= R^{A}\left(W\right) J_{A}+R^{A}\left(E\right) P_{A}\,,\label{F2f}
\end{equation}
where
\begin{eqnarray}
 R^{A}\left(W\right)&=&dW^{A}+\frac{1}{2}\epsilon^{ABC}W_{B}W_{C} \,,\notag\\
 R^{A}\left(E\right)&=& T^{A}-\frac{1}{\ell}\epsilon^{ABC} E_{B}E_{C} \,.\label{curvatures}
\end{eqnarray}
Here $R^{A}$ is the Lorentz curvature two-form and $T^{A}$ denotes the usual torsion two-form $T^{A}=dE^{A}+\epsilon^{ABC}W_{B}E_{C}$. The CS gravity action invariant under the teleparallel algebra \eqref{telea} can be obtained considering the one-form gauge connection $A$ \eqref{A1f} and the non-vanishing components of the invariant bilinear form \eqref{IT1} in the general expression for the CS action:
\begin{equation}\label{CS-Action}
    I_{\text{CS}}[A]= \frac{k}{4 \pi} \int_{\mathcal{M}} \langle A d A+ \frac{2}{3} A^{3}  \rangle \,.
\end{equation}
Then, the teleparallel CS action reads
\begin{eqnarray}
    I_{\text{TG}} &=& \frac{k}{4 \pi } \int_{\mathcal{M}} \bigg\{ \alpha_0 \left( W^A d W_A + \frac{1}{3}\epsilon^{ABC}W_A W_B W_C
    \right)  \notag\\
    &&+ \alpha_1 \left( 2 E_{A}R^{A}\left(W\right) + \frac{4}{3\ell^{2}} \epsilon^{ABC} E_A E_B E_C - \frac{2}{\ell}T^A E_A \right)\bigg\}\,, \label{TG}
\end{eqnarray}
where $k$ is the CS coupling constant and is related to the gravitational constant $G$ as $k=1/\left(4\pi G\right)$. The teleparallel CS action is characterized by the Einstein-Hilbert term, the comoslogical constant term plus a torsional term along the $\alpha_1$ constant. On the other hand, the sector proportional to $\alpha_0$ is known as the exotic CS term \cite{Witten:1988hc} and, as was mentioned in \cite{Caroca:2021njq}, it can be made to vanish without loss of generality. Nevertheless, for $\alpha_0\neq0$, the non-degeneracy of the invariant tensor \eqref{IT1} requires $\alpha_0\neq -\frac{\ell }{2}\alpha_1$ and $\alpha_1\neq0$. One can then see that, for non-degenerate invariant tensor, the equation of motions are given by the vanishing  of the curvature two-form \eqref{curvatures}. Unlike AdS gravity, which describes a Riemannian spacetime, the teleparallel CS gravity is characterized by a non-Riemannian geometry\footnote{See, e.g., \cite{Klemm:2018bil} for a recent review on non-Riemannian structures and applications in (super)gravity. Besides, we refer the reader to, e.g., \cite{Klemm:2019izb,Iosifidis:2021iuw,DAuria:2021dth,Myrzakulov:2021vel,Iosifidis:2021fnq} for further implications of a non-vanishing torsion and non-Riemannian structures in gravity and supergravity theories.} in which the cosmological constant can be seen as a source for the torsion,
\begin{equation}
T^{A}-\frac{1}{\ell}\epsilon^{ABC} E_{B}E_{C} = 0\,.\label{EOM}
\end{equation}
In fact, the latter is an on-shell constraint for the theory. In the vanishing cosmological constant limit $\ell\rightarrow\infty$, the theory reduces to the usual Poincaré CS gravity theory with vanishing torsion.

Let us note that the teleparallel gravity action can be recovered alternatively as a particular case of the Mielke-Baelker gravity model \cite{Mielke:1991nn,Baekler:1992ab,Cacciatori:2005wz}, whose action reads
\begin{equation}\label{MB}
    I_{\text{MB}}=aI_{1}+\Lambda I_{2}+\beta_{3} I_{3}+\beta_{4} I_{4} \,,
\end{equation}
where $a,\Lambda,\beta_{3}$ and $\beta_{4}$ are constants and
\begin{eqnarray}
I_{1} & = & 2\int E_{A}R^{A}\left(W\right)\,,\notag\\
I_{2} & = & -\frac{1}{3}\int \epsilon_{ABC}E^{A}E^{B}E^{C}\,,\notag\\
I_{3} & = & \int W^A dW_A + \frac{1}{3}\epsilon^{ABC}W_A W_B W_C\,,\notag\\
I_{4} & = &\int E_{A}T^{A}\,.\label{MBterms}
\end{eqnarray}
Indeed, by fixing the Mielke-Baelker parameters as follows,
\begin{equation}\label{MBcts}
    a=\frac{\alpha_1}{16\pi G}\,, \qquad \Lambda=-\frac{\alpha_1}{4\pi G \ell^2}\,, \qquad
    \beta_{3}= \frac{\alpha_0}{16 \pi G}\,, \qquad \beta_{4}=-\frac{\alpha_1}{8\pi G \ell}\,,
\end{equation}
we recover the teleparallel CS gravity action \eqref{TG} for $k=1/\left(4\pi G\right)$.


\section{Non-relativistic gravity with torsion}\label{sec3}

In this section we present the explicit construction of a non-relativistic version of the teleparallel CS gravity theory previously discussed. The NR action obtained is characterized by vanishing curvatures for the NR analogue of the spin connection and non-vanishing spatial torsion. Interestingly, the spatial component of the torsion is zero in the vanishing cosmological constant limit $\ell\rightarrow\infty$ and the NR gravity theory reproduces the known extended Bargmann gravity. The novel NR gravity theory presented here will be denoted as torsional NR gravity and offers us an alternative way to introduce a cosmological constant into the three-dimensional extended Bargmann gravity model.

\subsection{$U\left(1\right)$-enlargements}

A torsional NR algebra can be obtained by applying the NR limit to a particular $U\left(1\right)$-enlargement of the teleparallel algebra \eqref{telea}. To this end, let us introduce two extra $U\left(1\right)$ gauge fields to the field content. Then, the gauge connection one-form reads
\begin{equation}
    A= W^{A}J_{A}+E^{A} P_{A}+M Y_{1}+S Y_{2}\,.\label{A1fb} 
\end{equation}
As we shall see, the presence of these additional Abelian generators is required to ensure a finite and non-degenerate NR action in the limit process. Such extra gauge generators yield the following non-vanishing components of the invariant tensor:
\begin{equation}
    \langle Y_{2} Y_{2} \rangle = \alpha_0 \,,\qquad \quad \langle Y_{1} Y_{2} \rangle = \alpha_1 \,,\qquad \quad \langle Y_{1} Y_{1} \rangle = -\frac{2\alpha_1}{\ell} \,.\label{IT2}
\end{equation}
Then, the [teleparallel]$\oplus \mathfrak{u}\left(1\right)^{2}$ algebra admits the non-vanishing components of the invariant tensor given by \eqref{IT1} along with \eqref{IT2}. One can show that the relativistic CS action based on the [teleparallel]$\oplus \mathfrak{u}\left(1\right)^{2}$ algebra is obtained considering the gauge connection one-form \eqref{A1fb} and the invariant tensor \eqref{IT1} and \eqref{IT2} in the general expression of the CS action \eqref{CS-Action},
\begin{eqnarray}
    I_{\text{R}} &=& \frac{k}{4 \pi} \int_{\mathcal{M}} \bigg\{ \alpha_0 \left( W^A d W_A + \frac{1}{3}\epsilon^{ABC}W_A W_B W_C +SdS
    \right) \notag \\ 
    &+& \alpha_1 \left( 2 E_{A}R^{A}\left(W\right) + \frac{4}{3\ell^{2}} \epsilon^{ABC} E_A E_B E_C - \frac{2}{\ell}T^A E_A + 2MdS -\frac{2}{\ell}MdM \right)\bigg\}\,.\label{TU1G}
\end{eqnarray}
Let us note that we recover the relativistic CS action for the [Poincaré]$\oplus \mathfrak{u}\left(1\right)^2$ symmetry in the flat limit $\ell\rightarrow\infty$.

\subsection{Torsional non-relativistic gravity theory}

The NR counterpart of the relativistic [teleparallel]$\oplus \mathfrak{u}\left(1\right)^{2}$ algebra can be derived through an Inönü-Wigner procedure \cite{Inonu:1953sp}. To this end, we shall first consider the decomposition of the $A$ index as follows:
\begin{equation}
    A\rightarrow\left(0,a\right), \qquad  a=1,2\,. \label{decom}
\end{equation}
Then, we will introduce a dimensionless parameter $\xi$ which will allows us to identify the relativistic generators in terms of the NR ones (denoted with a tilde) as:
\begin{eqnarray}
J_{0} &=&\frac{\tilde{J}}{2}+\xi ^{2}\tilde{S}\,,\text{\ \ \ \ \ }J_{a}=\xi
\tilde{G}_{a}\,,\text{ \ \ \ \ \ \ }Y_{2}=\frac{\tilde{J}}{2}-\xi ^{2}\tilde{%
S}\,,  \notag \\
P_{0} &=&\frac{\tilde{H}}{2\xi }+\xi \tilde{M}\,,\text{ \ \ \ }P_{a}=\tilde{P%
}_{a}\,,\text{ \ \ \ \ \ \ \ \thinspace }Y_{1}=\frac{\tilde{H}}{2\xi }-\xi
\tilde{M}\,. \label{ident}
\end{eqnarray}
Furthermore, in order to ensure a finite NR symmetry after the contraction process we shall consider the following scaling of the length parameter:
\begin{equation}
    \ell\rightarrow \xi\ell\,.\label{ell}
\end{equation}
Then, considering the identification \eqref{ident} and the limit $\xi\rightarrow\infty$ in the relativistic commutation relations \eqref{telea}, we obtain a novel NR symmetry spanned by the set of generators $\{\tilde{J},\tilde{G}_{a},\tilde{S},\tilde{H},\tilde{P}_{a},\tilde{M}\}$ which satisfy the following commutators:
\begin{eqnarray}
\left[ \tilde{J},\tilde{G}_{a}\right] &=&\epsilon _{ab}\tilde{G}_{b}\,,\text{
\ \ \ \ \ \ \ \ }\left[ \tilde{G}_{a},\tilde{G}_{b}\right] =-\epsilon _{ab}%
\tilde{S}\,,\text{\ \ \ \ \ \ }\,\left[ \tilde{H},\tilde{G}_{a}\right]
=\epsilon _{ab}\tilde{P}_{b}\,,\text{ \ \ }  \notag \\
\left[ \tilde{J},\tilde{P}_{a}\right] &=&\epsilon _{ab}\tilde{P}%
_{b}\,,\qquad \quad \,\,\left[ \tilde{G}_{a},\tilde{P}_{b}\right] =-\epsilon
_{ab}\tilde{M}\,,\text{ \ \ \ \ \ }\left[ \tilde{H},\tilde{P}_{a}\right]
=-\frac{2}{\ell}\epsilon _{ab}\tilde{P}_{b}\,,  \notag \\
\left[ \tilde{P}_{a},\tilde{P}_{b}\right]
&=&\frac{2}{\ell}\epsilon _{ab}\tilde{M}\,, \label{torNR} 
\end{eqnarray}
where we have defined $\epsilon
^{ab}\equiv \epsilon ^{0ab}$ and $\epsilon _{ab}\equiv \epsilon _{0ab}$. As in the extended Newton-Hooke symmetry \cite{Papageorgiou:2010ud,Hartong:2016yrf}, the torsional NR algebra \eqref{torNR} consists of spatial
translations $\tilde{P}_a$, spatial rotations $\tilde{J}$, Galilean boosts $\tilde{G}_a$, time translations $\tilde{H}$ and two central charges $\tilde{S}$ and $\tilde{M}$. Let us note that the extended Bargmann algebra \cite{Bergshoeff:2016lwr} is obtained in the vanishing cosmological constant limit $\ell\rightarrow\infty$. On the other hand, the NR symmetry \eqref{torNR} can be written as two copies of the Nappi-Witten algebra \cite{Nappi:1993ie,Figueroa-OFarrill:1999cmq},
\begin{eqnarray}
\left[ \tilde{J}^{\pm},\tilde{G}^{\pm}_{a}\right]  &=&\epsilon _{ab}\tilde{G}^{\pm}_{b}\,,
\notag \\
\left[ \tilde{G}^{\pm}_{a},\tilde{G}^{\pm}_{b}\right]  &=&-\epsilon _{ab}\tilde{S}^{\pm}\,,
\label{NW}
\end{eqnarray}
by considering the following redefinitions:
\begin{eqnarray}
\tilde{G}_{a}&=&\tilde{G}^{+}_a + \tilde{G}^{-}_a\,, \qquad \qquad \ \  \tilde{P}_a=-\frac{2}{\ell}\tilde{G}^{-}_a\,, \notag \\
\tilde{J}&=&\tilde{J}^{+}+\tilde{J}^{-}\,, \qquad \qquad \ \ \ \,  \tilde{H}=-\frac{2}{\ell}\tilde{J}^{-} \,, \notag \\
\tilde{S}&=&\tilde{S}^{+}+\tilde{S}^{-}\,, \qquad \qquad \ \ \  \tilde{M}=-\frac{2}{\ell}\tilde{S}^{-} \,. \label{redef1}
\end{eqnarray}
Although the central charges can be set to zero, their presence is essential in the torsional NR algebra \eqref{torNR} in order to admit a non-degenerate invariant bilinear trace:
\begin{eqnarray}
\left\langle \tilde{J}\tilde{S}\right\rangle &=&-\tilde{\alpha}_{0}\,,\text{
\ \ \ \ \ \ \ \ \ \ \ \ \ \ \ \ \ \ \ \ \ \ \ \ \ \ }  \notag \\
\left\langle \tilde{G}_{a}\tilde{G}_{b}\right\rangle &=&\tilde{\alpha}%
_{0}\delta _{ab}\,,  \notag \\
\left\langle \tilde{G}_{a}\tilde{P}_{b}\right\rangle &=&\tilde{\alpha}%
_{1}\delta _{ab}\,,  \notag \\ 
\left\langle \tilde{H}\tilde{S}\right\rangle &=&\left\langle \tilde{M}\tilde{%
J}\right\rangle =-\tilde{\alpha}_{1}\,,  \notag \\
\left\langle \tilde{P}_{a}\tilde{P}_{b}\right\rangle &=& =-\frac{2\tilde{\alpha}_1}{\ell}\delta _{ab}\,,
\notag \\
\left\langle \tilde{H}\tilde{%
M}\right\rangle &=&\frac{2\tilde{\alpha%
}_{1}}{\ell}\,,  \label{torIT}
\end{eqnarray}%
where the relativistic parameters $\alpha $'s have been rescaled as
\begin{equation}
\alpha _{0}=\tilde{\alpha}_{0}\xi ^{2}\,, \qquad \qquad \alpha _{1}=\tilde{%
\alpha}_{1}\xi \,.  \label{alphas}
\end{equation}
Analogously to \cite{Aviles:2018jzw,Concha:2019lhn,Concha:2020sjt}, such rescaling is required in order to have a finite NR
CS action after the IW contraction. One can see that the non-degeneracy of the invariant tensor is preserved as long as $\tilde{\alpha}_0\neq-\frac{\ell}{2}\tilde{\alpha_1}$ and $\tilde{\alpha}_1\neq 0$. The non-degeneracy of the invariant tensor is related to the requirement that the NR CS gravity action involves a kinematical term for each gauge field. In particular, the corresponding gauge connection one-form for the torsional NR algebra \eqref{torNR} reads
\begin{equation}
\tilde{A} =\omega \tilde{J}+\omega ^{a}%
\tilde{G}_{a}+\tau \tilde{H}+e^{a}\tilde{P}_{a}+m\tilde{M}+s\tilde{S}\,.
\label{tor1f}
\end{equation}
The curvature associated to this gauge connection is given by
\begin{eqnarray}
\tilde{F} &=&R(\omega )\tilde{J}%
+R^{a}(\omega ^{b})\tilde{G}_{a}+R(\tau )\tilde{H}+R^{a}(e^{b})\tilde{P}_{a}+R(m)%
\tilde{M}+R(s)\tilde{S}\,,\label{torcuv}
\end{eqnarray}%
where
\begin{eqnarray}
R(\omega ) &=&d\omega \,,  \notag \\
R^{a}(\omega ^{b}) &=&d\omega ^{a}+\epsilon ^{ac}\omega \omega _{c}\,,
\notag \\
R(\tau ) &=&d\tau \,,  \notag \\
R^{a}(e^{b}) &=&de^{a}+\epsilon ^{ac}\omega e_{c}+\epsilon ^{ac}\tau \omega
_{c}-\frac{2}{\ell}\epsilon^{ac}\tau e_{c}\,,  \notag \\
R(m) &=&dm+\epsilon ^{ac}e_{a}\omega _{c}-\frac{1}{\ell}\epsilon^{ac}e_{a}e_{c}\,,  \notag \\
R(s) &=&ds+\frac{1}{2}\epsilon ^{ac}\omega _{a}\omega _{c}\,. \label{curvaturesNR}
\end{eqnarray}%
The respective CS action based on the torsional NR algebra \eqref{torNR} is obtained considering the non-vanishing components of the invariant tensor \eqref{torIT} and the gauge connection one-form \eqref{tor1f} in the general expression of the CS action \eqref{CS-Action}:
\begin{eqnarray}
I_{\text{NR}} &=&\frac{k}{4\pi}\int \tilde{\alpha}_{0}\left[ \omega _{a}R^{a}(\omega
^{b})-2sR\left( \omega \right) \right] +\tilde{\alpha}_{1}\left[
2e_{a}R^{a}(\omega ^{b})-2mR(\omega )-2\tau R(s)\right. \notag \\
&&-\left.\frac{2}{\ell}e_{a}R^{a}\left( e^{b}\right)+\frac{2}{\ell}mR\left( \tau \right)+\frac{2}{\ell}\tau R\left(m\right)+\frac{2}{\ell^2}\tau\epsilon^{ac}e_{a}e_{c}\right]  \,.  \label{torCS}
\end{eqnarray}%
The NR CS action \eqref{torCS} is the most general NR CS action invariant under the torsional NR algebra \eqref{torNR}. One can notice that it contains two independent sectors and can be seen as the NR version of the three-dimensional teleparallel CS gravity theory presented in \cite{Caroca:2021njq}. The first term is indistinguishable from the exotic sector of the extended Bargmmann gravity \cite{Concha:2020eam}, being the corresponding NR version of the exotic gravity \cite{Witten:1988hc}. The term proportional to $\tilde{\alpha}_{1}$ can be seen as a cosmological extension of the extended Bargmann gravity action \cite{Bergshoeff:2016lwr}. However, the inclusion of the cosmological constant through the torsional NR algebra is quite different from the one constructed using the extended Newton-Hooke symmetry. Indeed, the field equations derived from \eqref{torCS} are given by the vanishing of the curvature two-forms \eqref{curvaturesNR}. One can see that the field equations for $\omega^{a}$, $e^{a}$, $s$ and $m$ are diverse from the Newton-Hooke ones, the latter being given by
\begin{eqnarray}
F^{a}(\omega ^{b}) &=&d\omega ^{a}+\epsilon ^{ac}\omega \omega _{c}+\frac{1}{\ell^{2}}\epsilon^{ac}\tau e_c=0\,, \notag \\
F^{a}(e^{b}) &=&de^{a}+\epsilon ^{ac}\omega e_{c}+\epsilon ^{ac}\tau \omega
_{c}=0\,, \notag\\
F(s) &=&ds+\frac{1}{2}\epsilon ^{ac}\omega _{a}\omega _{c}+\frac{1}
{2\ell^2}\epsilon^{ac}e_{a}e_{c}=0\,, \notag \\
F(m) &=&dm+\epsilon ^{ac}e_{a}\omega _{c}=0\,. 
\end{eqnarray}
In our torsional NR gravity theory, the cosmological constant can be seen as a source for the spatial torsion $T^{a}\left(e^{b}\right)=de^{a}+\epsilon ^{ac}\omega e_{c}+\epsilon ^{ac}\tau \omega_{c}$ and for the curvature $T\left(m\right)=dm+\epsilon ^{ac}e_{a}\omega _{c}$. Indeed, on-shell we find
\begin{eqnarray}
T^{a}\left(e^{b}\right)&=&\frac{2}{\ell}\epsilon^{ac}\tau e_{c}\,,\notag \\
T\left(m\right)&=&\frac{1}{\ell}\epsilon^{ac}e_{a}e_{c}\,.
\end{eqnarray} 
The NR counterpart of the teleparallel CS gravity \cite{Caroca:2021njq} is then described, on-shell, by vanishing curvatures $R\left(\omega\right)=0=R^{a}\left(\omega^{a}\right)$ and a non-vanishing spatial torsion $T^{a}\left(e^{b}\right)\neq0$. Naturally, we recover the extended Bargmann gravity theory \cite{Bergshoeff:2016lwr} with a vanishing spatial torsion in the vanishing cosmological constant $\ell\rightarrow\infty$.

Let us note that the NR gravity action \eqref{torCS} can be alternatively recovered from the relativistic $U\left(1\right)$-enlarged teleparallel CS action \eqref{TU1G}. Indeed, one can express the relativistic gauge fields in terms of the NR ones as follows:
\begin{eqnarray}
W^{0} &=&\omega +\frac{s}{2\xi ^{2}}\,,\text{\ \ \ \ \ }W^{a}=\frac{\omega
^{a}}{\xi }\,,\text{ \ \ \ \ \ \ \thinspace \thinspace \thinspace }S=\omega -%
\frac{s}{2\xi ^{2}}\,,  \notag \\
E^{0} &=&\xi \tau +\frac{m}{2\xi }\,,\text{ \ \ \ \ \thinspace \thinspace }%
E^{a}=e^{a}\,,\text{ \ \ \ \ \ \ \ \ }M=\xi \tau -\frac{m}{2\xi }\,, \label{rescgf}
\end{eqnarray}
which satisfy $A=\tilde{A}$. The NR CS action \eqref{torCS} appears considering these last expressions along the rescaling of the relativistic parameters \eqref{alphas} on the relativistic CS action \eqref{TU1G} and then applying the limit $\xi\rightarrow\infty$.


\section{Non-relativistic supergravity with torsion}\label{sec4}

We will now present the explicit supersymmetric extension of the torsional NR gravity theory previously discussed. The NR action will be characterized, on-shell, by the vanishing of the NR curvatures but non-vanishing spatial super-torsion.

A torsional NR superalgebra can be obtained by applying the S-expansion method to the $\mathcal{N}=2$ teleparallel superalgebra. The S-expansion procedure will also provide us with a non-degenerate invariant tensor and therefore we will be able to produce a well-defined torsional NR CS supergravity action which can be seen as the corresponding NR counterpart of the $\mathcal{N}=2$ teleparallel supergravity action presented in \cite{Caroca:2021njq}.

\subsection{Torsional non-relativistic superalgebra}

We shall now move on to the supersymmetric extension of the results presented in the previous section. To this aim, we will focus on the relativistic $\mathcal{N}=2$ teleparallel superalgebra presented in \cite{Caroca:2021njq}, and study its non-relativistic expansion.

At the relativistic level, $\mathcal{N}$-extended teleparallel supergravity has been formulated as a CS theory invariant under the so-called $\mathcal{N}$-extended teleparallel superalgebra \cite{Caroca:2021njq}. In the $\mathcal{N}=2$ case, which is the one of interest in the present paper, the teleparallel superalgebra is spanned by the bosonic generators $(J_A,P_A)$ along with the internal symmetry generator $Z$, and the fermionic charges $Q^i_\alpha$, with $i=1,2$ and $\alpha=1,2$, which are two-components Majorana spinor charges.\footnote{For simplicity, in the following we will frequently omit writing of the spinor indices.} However, in order to have a well-defined flat limit $\ell \rightarrow \infty$ leading to the $\mathcal{N}=2$ Poincaré superalgebra extended with an $\mathfrak{so}(2)$ automorphism algebra \cite{Howe:1995zm}, the $\mathcal{N}=2$ teleparallel superalgebra was extended in \cite{Caroca:2021njq} to include the automorphism generator $S$. To this end, the redefinition $\mathcal{T}=Z-\frac{\ell}{2}\mathcal{S}$ was performed, obtaining the following non-trivial (anti-)commutation relations in the $\mathcal{N}=2$ case:
\begin{eqnarray}
\left[J_A,Q_{\alpha}^{i}\right]&=&-\frac{1}{2}\left(\gamma_A\right)_{\alpha}^{\ \beta}Q_{\beta}^{i}\,, \notag \\
\left[\mathcal{T},Q_{\alpha}^i\right]&=&-\epsilon^{ij} Q^j_\alpha \,, \notag \\
\left\{Q_{\alpha}^{i},Q_{\beta}^{j}\right\}&=&-\delta^{ij}\left(\gamma^AC\right)_{\alpha\beta}\left(P_A+\frac{2}{\ell}J_A\right)+C_{\alpha\beta} \epsilon^{ij} \left(\frac{2}{\ell}\mathcal{T}+\mathcal{S}\right) \,,\label{superteleNtwo} 
\end{eqnarray}
along with \eqref{telea}. The gamma matrices in three dimensions are denoted by $\gamma_A$ and $C$ is the charge conjugation matrix, satisfying $C^{T}=-C$ and $C\gamma^{A}=(C\gamma^{A})^{T}$, while $\epsilon^{ij}$ is the rank-2 Levi-Civita symbol ($\epsilon^{12}=1$, $\epsilon^{21}=-1$). 
The $\mathcal{N}=2$ Poincaré superalgebra extended with the $\mathfrak{so}(2)$ automorphism algebra is properly recovered in the flat limit $\ell \rightarrow \infty$. The presence of the automorphism generator in the Poincaré case is required in order to define a non-degenerate invariant tensor \cite{Howe:1995zm}. Observe that, as in the case of the $\mathcal{N}=2$ Poincaré superalgebra, the generator $\mathcal{S}$ is a central charge. Even though the $\mathcal{N}=2$ teleparallel superalgebra given by \eqref{telea} and \eqref{superteleNtwo} presents a well-defined Poincaré limit, its (anti-)commutation relations are quite different from the super-AdS ones.
Let us stress that the $\mathcal{N}=2$ teleparallel superalgebra can be written as the direct sum of $\mathfrak{osp}(2|2)\otimes \mathfrak{sp}(2)$ and the $\mathfrak{so}(2)$ automorphism algebra by considering the following identification of the generators:
\begin{equation}\label{redefospaut}
    L_{A}\equiv J_{A} +\frac{\ell}{2} P_{A} \,, \quad S_A\equiv -\frac{\ell}{2} P_{A}\,, \quad \mathcal{G}_{\alpha}^{i}\equiv \sqrt{\frac{\ell}{2}}Q_{\alpha}^{i}\,, \quad \mathcal{M}\equiv \mathcal{T}+\frac{\ell}{2}\mathcal{S} \,, \quad B\equiv -\frac{\ell}{2}\mathcal{S}\,, 
\end{equation}
where $(L_A,S_A,\mathcal{M},\mathcal{G}_{\alpha}^{i})$ satisfy the $\mathfrak{osp}(2|2)\otimes \mathfrak{sp}(2)$ superalgebra, while $B$ is the $\mathfrak{so}(2)$ automorphism generator.

The $\mathcal{N}=2$ teleparallel superalgebra is endowed with the following non-vanishing components of the (non-degenerate) invariant tensor:
\begin{eqnarray}
    \langle \mathcal{T} \mathcal{T} \rangle &=& - 2\alpha_0 \,, \notag \\
    \langle \mathcal{T} \mathcal{S} \rangle &=& - 2\alpha_1 \,, \notag \\
      \langle \mathcal{S} \mathcal{S} \rangle &=& \frac{4\alpha_1}{\ell} \,, \notag \\
    \langle Q_{\alpha}^{i} Q_{\beta}^{j}\rangle &=& 2\left(\frac{2\alpha_0}{\ell}+\alpha_1\right)C_{\alpha\beta}\delta^{ij}\,, \label{IT3}
\end{eqnarray}
along with \eqref{IT1}. We refer the reader to \cite{Caroca:2021njq} for the way in which the constant parameters $\alpha_0$ and $\alpha_1$ are related to the $\mathfrak{osp}(2|2)\otimes \mathfrak{sp}(2)$ and $\mathfrak{so}(2)$ ones. The flat limit $\ell \rightarrow \infty$ yields the invariant tensor for the $\mathcal{N}=2$ Poincaré superalgebra extended with $\mathfrak{so}(2)$ \cite{Howe:1995zm}.

A NR teleparallel superalgebra can be obtained from the $\mathcal{N}=2$ relativistic one by performing an S-expansion \cite{Izaurieta:2006zz}. Indeed, as it was also seen in \cite{Gomis:2019nih, Concha:2020tqx,Concha:2021jos}, the S-expansion of a relativistic superalgebra considering $S^{(2)}_E$ as the relevant semigroup reproduces a NR contraction. This is due to the fact that the S-expansion method can be seen as a generalization of the Inönü-Wigner contraction process \cite{Inonu:1953sp} when the semigroup considered belongs to the $S^{(N)}_E$ family. Here we focus on the semigroup $S^{(2)}_E$, which allows to obtain not only a well-defined NR teleparallel superalgebra but also its non-degenerate invariant tensor.

We first proceed as in Section \ref{sec3} by considering the decomposition of the $A$ index given in \eqref{decom}. Besides, we perform a redefinition of the supercharges by defining
\begin{equation}
    Q^{\pm}_\alpha = \frac{1}{\sqrt{2}} \left( Q^1_\alpha \pm \left(\gamma^0\right)_{\alpha \beta} Q^2_\beta \right) \,.
\end{equation}
Then, we consider $S_{E}^{(2)}=\{ \lambda_0,\lambda_1,\lambda_2,\lambda_3 \}$ as the relevant semigroup to perform the expansion of the relativistic $\mathcal{N}=2$ teleparallel superalgebra. The elements of $S_{E}^{(2)}$ satisfy the following multiplication law:
\begin{equation}
\begin{tabular}{l|llll}
$\lambda _{3}$ & $\lambda _{3}$ & $\lambda _{3}$ & $\lambda _{3}$ & $\lambda
_{3}$ \\
$\lambda _{2}$ & $\lambda _{2}$ & $\lambda _{3}$ & $\lambda _{3}$ & $\lambda
_{3}$ \\
$\lambda _{1}$ & $\lambda _{1}$ & $\lambda _{2}$ & $\lambda _{3}$ & $\lambda
_{3}$ \\
$\lambda _{0}$ & $\lambda _{0}$ & $\lambda _{1}$ & $\lambda _{2}$ & $\lambda
_{3}$ \\ \hline
& $\lambda _{0}$ & $\lambda _{1}$ & $\lambda _{2}$ & $\lambda _{3}$%
\end{tabular}
\label{mlStwoE}
\end{equation}
where $\lambda_3=0_s$ is the zero element of the semigroup such that $0_s\lambda_k=0_s$, $k = 1,2,3$. Before applying the S-expansion procedure, we consider a particular subspace decomposition of the $\mathcal{N}=2$ teleparallel superalgebra. To this end, let us first consider the split of the index $A$ along with the redefinition of the supercharges above.
Hence, we can see that the subspaces decomposition of the teleparallel superalgebra given by $V_0 = \lbrace J_0, P_0, \mathcal{T}, \mathcal{S}, Q^+_\alpha \rbrace$ and $V_1 = \lbrace J_a, P_a, Q^-_\alpha \rbrace$ satisfies
\begin{eqnarray}
[V_0,V_0]\subset V_0\,,\qquad [V_0,V_1]\subset V_1\,, \qquad [V_1,V_1]\subset V_0\,.
\end{eqnarray}
Let us consider now $S_{E}^{(2)}=S_{0} \cup S_{1}$ as decomposition of the relevant semigroup $S_{E}^{(2)}$, where
\begin{eqnarray}
S_0&=&\{\lambda_0,\lambda_2,\lambda_3\}\,, \notag \\
S_1&=&\{\lambda_1,\lambda_3\} \,. 
\label{decomp}
\end{eqnarray}
Then, the decomposition \eqref{decomp} is said to be resonant since it satisfies the same structure as the subspaces, that is
\begin{equation}
    S_0\cdot S_0\subset S_0\,,\qquad \ S_0\cdot S_1\subset S_1\,,\qquad \ S_1\cdot S_1\subset S_0\,.\label{semidecomp}
\end{equation}
Following \cite{Izaurieta:2006zz}, after extracting a resonant subalgebra of the $S_{E}^{(2)}$-expansion of the $\mathcal{N}=2$ teleparallel superalgebra and applying a $0_s$-reduction, one ends up with a new NR expanded superalgebra spanned by the set of generators
\begin{equation}
\{\tilde{J},\tilde{G}_{a},\tilde{S},\tilde{H},\tilde{P}_{a},\tilde{M},\tilde{T}_1,\tilde{T}_2,\tilde{U}_1,\tilde{U}_2,\tilde{Q}_{\alpha }^{+},\tilde{R}_{\alpha},\tilde{Q}_{\alpha
}^{-}\}\,,
\end{equation}
which are related to the relativistic ones through the semigroup elements as
\begin{equation}
    \begin{tabular}{lll}
\multicolumn{1}{l|}{$\lambda_3$} & \multicolumn{1}{|l}{\cellcolor[gray]{0.8}} & \multicolumn{1}{|l|}{\cellcolor[gray]{0.8}} \\ \hline
\multicolumn{1}{l|}{$\lambda_2$} & \multicolumn{1}{|l}{$\tilde{S},\ \tilde{M},\ \tilde{T}_2,\,\tilde{U}_2,\,\tilde{R}_{\alpha}$} & \multicolumn{1}{|l|}{\cellcolor[gray]{0.8}} \\ \hline
\multicolumn{1}{l|}{$\lambda_1$} & \multicolumn{1}{|l}{\cellcolor[gray]{0.8}} & \multicolumn{1}{|l|}{$\tilde{G}_a,\,\tilde{P}_a,\,\tilde{Q}^{-}_{\alpha}$} \\ \hline
\multicolumn{1}{l|}{$\lambda_0$} & \multicolumn{1}{|l}{$ \tilde{J},\,\ \tilde{H},\ \tilde{T}_1,\,\tilde{U}_1,\,\tilde{Q}^{+}_{\alpha}$} & \multicolumn{1}{|l|}{\cellcolor[gray]{0.8}} \\ \hline
\multicolumn{1}{l|}{} & \multicolumn{1}{|l}{$J_0,\,P_0,\,\mathcal{T},\ \mathcal{S},\ \, Q^{+}_{\alpha}$} & \multicolumn{1}{|l|}{$J_{a},\ P_{a},\,Q^{-}_{\alpha}$} 
\end{tabular}\label{Sexp}%
\end{equation}
The NR generators satisfy precisely the purely bosonic subalgebra \eqref{torNR} along with the following (anti-)commutation relations:
\begin{eqnarray}
\left[ \tilde{J},\tilde{Q}_{\alpha }^{\pm }\right] &=&-\frac{1}{2}\left(
\gamma _{0}\right) _{\alpha }^{\text{ }\beta }\tilde{Q}_{\beta }^{\pm
}\,,\qquad \left[ \tilde{J},\tilde{R}_{\alpha }\right] =-
\frac{1}{2}\left( \gamma _{0}\right) _{\alpha }^{\text{ }\beta }\tilde{R}_{\beta }\,, \qquad \left[ \tilde{S},\tilde{Q}_{\alpha }^{+}\right] =-\frac{1}{2}\left(
\gamma _{0}\right) _{\alpha }^{\text{ }\beta }\tilde{R}_{\beta }\,, \notag\\
\left[ \tilde{G}_{a},\tilde{Q}_{\alpha }^{+}\right] &=&-\frac{1}{2}\left(
\gamma _{a}\right) _{\alpha }^{\text{ }\beta }\tilde{Q}_{\beta
}^{-}\,, \quad \ 
\left[ \tilde{G}_{a},\tilde{Q}_{\alpha }^{-}\right] =-\frac{
1}{2}\left( \gamma _{a}\right) _{\alpha }^{\text{ }\beta }\tilde{R}_{\beta
}\,, \notag \\
\left[ \tilde{T}_1,\tilde{Q}_{\alpha }^{\pm }\right] &=& \pm \left(\gamma ^{0}\right) _{\alpha \beta}\tilde{Q}_{\beta }^{\pm}\,,\qquad \left[ \tilde{T}_2,\tilde{Q}_{\alpha }^{+}\right] = \left(\gamma ^{0}\right) _{\alpha \beta}\tilde{R}_{\beta } \,, \qquad \left[ \tilde{T}_1,\tilde{R}_{\alpha }\right] = \left(\gamma ^{0}\right) _{\alpha \beta}\tilde{R}_{\beta } \,, \notag \\
\left\{ \tilde{Q}_{\alpha }^{+},\tilde{Q}_{\beta }^{+}\right\} &=&-\left(
\gamma ^{0}C\right) _{\alpha \beta } \left(\tilde{H}+\frac{2}{\ell} \tilde{J} \right) - \left( \gamma
^{0}C\right) _{\alpha \beta }\left(\frac{2}{\ell}\tilde{T}_1+\tilde{U}_1 \right)\,, \notag \\
\left\{ \tilde{Q}_{\alpha }^{-},\tilde{Q}_{\beta }^{-}\right\} &=&-\left(
\gamma ^{0}C\right) _{\alpha \beta } \left(\tilde{M}+\frac{2}{\ell} \tilde{S} \right) + \left( \gamma
^{0}C\right) _{\alpha \beta }\left(\frac{2}{\ell}\tilde{T}_2+\tilde{U}_2 \right)\,, \notag \\
\left\{ \tilde{Q}_{\alpha }^{+},\tilde{R}_{\beta }\right\} &=&-\left(
\gamma ^{0}C\right) _{\alpha \beta } \left(\tilde{M}+\frac{2}{\ell} \tilde{S} \right) - \left( \gamma
^{0}C\right) _{\alpha \beta }\left(\frac{2}{\ell}\tilde{T}_2+\tilde{U}_2 \right)\,, \notag \\
\left\{ \tilde{Q}_{\alpha }^{+},\tilde{Q}_{\beta }^{-}\right\} &=&-\left(
\gamma ^{a}C\right) _{\alpha \beta } \left(\tilde{P}_a + \frac{2}{\ell} \tilde{G}_a \right) \,. \label{superNRtelpart}
\end{eqnarray}
The superalgebra given by \eqref{torNR} and \eqref{superNRtelpart} can be seen as the supersymmetric extension of the torsional NR algebra obtained previously \eqref{torNR}. One can see that such superalgebra requires the introduction of four additional bosonic generators $\tilde{T}_1$, $\tilde{T}_2$, $\tilde{U}_1$ and $\tilde{U}_2$. Notice that the generators $\tilde{U}_1$ and $\tilde{U}_2$ are central charges, while $\tilde{T}_1$ and $\tilde{T}_2$\, which arise from the S-expansion of the relativistic R-symmetry generator $\mathcal{T}$, act non-trivially on the spinor charges $\tilde{Q}_{\alpha}^{\pm}$ and $\tilde{R}_{\alpha}$. Although the central charges can be set to zero by performing a simple contraction, their presence here is essential in order to have a non-degenerate invariant tensor. Interestingly, we recover the extended Bargmann superalgebra \cite{Bergshoeff:2016lwr} in the flat limit $\ell\rightarrow\infty$ and considering $\tilde{T}_1=\tilde{T}_2=\tilde{U}_1=\tilde{U}_2=0$. Such behavior is inherited from the respective relativistic counterparts, in which case the teleparallel algebra reproduces the Poincaré algebra in the vanishing cosmological constant limit. The inclusion of a cosmological constant into the extended Bargmann supergravity has already been approached considering a supersymmetric extension of the extended Newtoon-Hooke symmetry \cite{Concha:2020eam}. Here we present an alternative NR superalgebra allowing the presence of a cosmological constant, but with a non-vanishing spatial super-torsion as we shall see.

Before approaching the construction of a CS supergravity action based on the torsional NR superalgebra, let us stress that the NR superalgebra given by \eqref{torNR} and \eqref{superNRtelpart} can be written as two copies of the Nappi-Witten algebra, one of which is augmented by supersymmetry endowed with $\mathfrak{u}\left(1\right)$ generators. To this end, let us consider the following redefinition of the generators of the torsional NR superalgebra:
\begin{eqnarray}
\tilde{G}_{a}&=&G_a+\hat{G}_a\,, \qquad \qquad \ \  \tilde{P}_a=-\frac{2}{\ell}\hat{G}_a\,, \qquad \qquad \tilde{Q}_{\alpha}^{+}=\sqrt{\frac{2}{\ell}}\mathcal{Q}_{\alpha}^{+}\, \notag \\
\tilde{J}&=&J+\hat{J}\,, \qquad \qquad \qquad \tilde{H}=-\frac{2}{\ell}\hat{J} \,, \qquad \qquad \  \ \tilde{Q}_{\alpha}^{-}=\sqrt{\frac{2}{\ell}}\mathcal{Q}_{\alpha}^{-}\,, \notag \\
\tilde{S}&=&S+\hat{S}\,, \qquad \qquad \quad \ \ \,  \tilde{M}=-\frac{2}{\ell}\hat{S} \,, \qquad \qquad \ \ \tilde{R}_{\alpha}=\sqrt{\frac{2}{\ell}}\mathcal{R}_{\alpha}\,, \notag \\
\tilde{T}_{1}&=&X_1+\hat{X}_1\,, \qquad \qquad \ \  \tilde{U}_{1}=-\frac{2}{\ell}\hat{X}_1\,, \notag \\
\tilde{T}_{2}&=&X_2+\hat{X}_2\,, \qquad \qquad \ \  \tilde{U}_{2}=-\frac{2}{\ell}\hat{X}_2\,.
\label{redef12}
\end{eqnarray}
Here, the subset spanned by $\{J,G_{a},S,X_1,X_2,\mathcal{Q}_{\alpha}^{+},\mathcal{Q}_{\alpha}^{-},\mathcal{R}_{\alpha}\}$ satisfy a supersymmetric extension of the Nappi-Witten algebra \cite{Concha:2020tqx,Concha:2020eam}:
\begin{eqnarray}
\left[ J,G_{a}\right] &=&\epsilon _{ab}G_{b}\,,\qquad \qquad \ \ \ \quad \left[
G_{a},G_{b}\right] =-\epsilon _{ab}S\,,  \notag \\
\left[ J,\mathcal{Q}_{\alpha }^{\pm }\right] &=&-\frac{1}{2}\left( \gamma _{0}\right)
_{\alpha }^{\,\ \beta }\mathcal{Q}_{\beta }^{\pm }\,,\quad \ \ \ \ \ \ \left[
J,\mathcal{R}_{\alpha }\right] =-\frac{1}{2}\left( \gamma _{0}\right) _{\alpha }^{%
\text{ }\beta }\mathcal{R}_{\beta }\,,  \notag \\
\left[ G_{a},\mathcal{Q}_{\alpha }^{+}\right] &=&-\frac{1}{2}\left( \gamma _{a}\right)
_{\alpha }^{\,\ \beta }\mathcal{Q}_{\beta }^{-}\,,\quad \ \ \ \left[ G_{a},\mathcal{Q}_{\alpha
}^{-}\right] =-\frac{1}{2}\left( \gamma _{a}\right) _{\alpha }^{\text{ }%
\beta }\mathcal{R}_{\beta }\,,  \notag \\
\left[ S,\mathcal{Q}_{\alpha }^{+}\right] &=&-\frac{1}{2}\left( \gamma _{0}\right)
_{\alpha }^{\text{ }\beta }\mathcal{R}_{\beta }\,,\quad \ \quad \left[
X_{1},\mathcal{Q}_{\alpha }^{\pm}\right] = \pm \frac{1}{2}\left( \gamma _{0}\right) _{\alpha
\beta }\mathcal{Q}_{\beta }^{\pm}\,,  \notag \\
\left[ X_{2},\mathcal{Q}_{\alpha }^{+}%
\right] &=& \frac{1}{2}\left( \gamma _{0}\right) _{\alpha \beta }\mathcal{R}_{\beta } \,,\qquad \quad \, \left[ X_{1},\mathcal{R}_{\alpha }\right] = \frac{1}{2}\left( \gamma _{0}\right)
_{\alpha \beta }\mathcal{R}_{\beta } \,, \notag \\
\left\{ \mathcal{Q}_{\alpha }^{+},\mathcal{Q}_{\beta }^{-}\right\} &=&-\left( \gamma
^{a}C\right) _{\alpha \beta }G_{a}\,,  \notag \\
\left\{ \mathcal{Q}_{\alpha }^{+},\mathcal{Q}_{\beta }^{+}\right\} &=&-\left( \gamma
^{0}C\right) _{\alpha \beta }J-\left( \gamma ^{0}C\right) _{\alpha \beta
}X_{1}\,,\text{ \qquad }  \notag \\
\left\{ \mathcal{Q}_{\alpha }^{-},\mathcal{Q}_{\beta }^{-}\right\} &=&-\left( \gamma
^{0}C\right) _{\alpha \beta }S {+} \left( \gamma ^{0}C\right) _{\alpha \beta
}X_{2}\,,  \notag \\
\left\{ \mathcal{Q}_{\alpha }^{+},\mathcal{R}_{\beta }\right\} &=&-\left( \gamma ^{0}C\right)
_{\alpha \beta }S-\left( \gamma ^{0}C\right) _{\alpha \beta }X_{2}\,. \label{sNW}
\end{eqnarray}
On the other hand, the set of generators $\{\hat{J},\hat{G}_{a},\hat{S},\hat{X}_1,\hat{X}_2\}$ satisfy a $U\left(1\right)$-enlargement of the usual Nappi-Witten algebra \eqref{NW}.

\subsection{Torsional non-relativistic supergravity theory}

Let us construct the explicit NR CS supergravity action invariant under the torsional NR superalgebra \eqref{torNR} and \eqref{superNRtelpart}. Although the Nappi-Witten algebraic structure seems simpler, we are interested in exploring an alternative way to accommodate a cosmological constant into the extended Bargmann supergravity different from considering the extended Newton-Hooke one \cite{Ozdemir:2019tby}. As we shall see, the inclusion of a cosmological constant through the NR superlgebra \eqref{torNR} and \eqref{superNRtelpart} will imply the presence of a non-vanishing spatial super-torsion.

One can show that the non-vanishing components of an invariant tensor for the NR teleparallel superalgebra just obtained are given by \eqref{torIT} along with
\begin{eqnarray}
    \langle \tilde{T}_1 \tilde{T}_2 \rangle &=& - 2\tilde{\alpha}_0 \,, \notag \\
    \langle \tilde{T}_1 \tilde{U}_2 \rangle &=& \langle \tilde{T}_2 \tilde{U}_1 \rangle = - 2\tilde{\alpha}_1 \,, \notag \\
      \langle \tilde{U}_1 \tilde{U}_2 \rangle &=& \frac{4\tilde{\alpha}_1}{\ell} \,, \notag \\
    \langle Q_{\alpha}^{+} R_{\beta} \rangle &=& 2\left(\frac{2\tilde{\alpha}_0}{\ell}+\tilde{\alpha}_1\right)C_{\alpha\beta} \,, \notag \\
    \langle Q_{\alpha}^{-} Q_{\beta}^{-}\rangle &=& 2\left(\frac{2\tilde{\alpha}_0}{\ell}+\tilde{\alpha}_1\right)C_{\alpha\beta} \,. \label{supertorIT}
\end{eqnarray}
Here the invariant tensor can be obtained from the relativistic one \eqref{IT1} and \eqref{IT3} considering \eqref{Sexp} and defining the NR parameters in terms of the relativistic ones through the semigroup elements as
\begin{equation}
    \tilde{\alpha}_0 = \lambda_2 \alpha_0 \,, \quad \tilde{\alpha}_1 = \lambda_2 \alpha_1 \,.\label{exp1}
\end{equation}

The gauge connection one-form for the torsional NR superalgebra given by \eqref{torNR} and \eqref{supertorIT} reads as follows:
\begin{equation}
\tilde{A} =\omega \tilde{J}+\omega ^{a}%
\tilde{G}_{a}+\tau \tilde{H}+e^{a}\tilde{P}_{a}+m\tilde{M}+s\tilde{S} + t_1 \tilde{T}_1 + t_2 \tilde{T}_2 + u_1 \tilde{U}_1 + u_2 \tilde{U}_2 + \bar{\psi}^+ \tilde{Q}^+ + \bar{\psi}^- \tilde{Q}^- + \bar{\rho} \tilde{R} \,.
\label{supertor1f}
\end{equation}
The curvature two-form associated with the gauge connection above is given by
\begin{eqnarray}
\tilde{F} &=&F(\omega )\tilde{J}%
+F^{a}(\omega ^{b})\tilde{G}_{a}+F(\tau )\tilde{H}+F^{a}(e^{b})\tilde{P}_{a}+F(m)%
\tilde{M}+F(s)\tilde{S} \notag \\
& & + F(t_1) \tilde{T}_1 + F(t_2) \tilde{T}_2 + F(u_1) \tilde{U}_1 + F(u_2) \tilde{U}_2 + \nabla \bar{\psi}^+ \tilde{Q}^+ + \nabla \bar{\psi}^- \tilde{Q}^- + \nabla \bar{\rho} \tilde{R} \,,\label{supertorcuv}
\end{eqnarray}%
where the bosonic curvatures are given by
\begin{eqnarray}
F(\omega ) &=& R(\omega) + \frac{1}{\ell} \bar{\psi}^+ \gamma^0 \psi^+ = d \omega + \frac{1}{\ell} \bar{\psi}^+ \gamma^0 \psi^+ \,,  \notag \\
F^{a}(\omega ^{b}) &=& R^{a}(\omega ^{b}) + \frac{2}{\ell} \bar{\psi}^+ \gamma^a \psi^- = d\omega ^{a}+\epsilon ^{ac}\omega \omega _{c} + \frac{2}{\ell} \bar{\psi}^+ \gamma^a \psi^- \,, \notag \\
F(\tau ) &=& R(\tau ) + \frac{1}{2} \bar{\psi}^+ \gamma^0 \psi^+ = d \tau + \frac{1}{2} \bar{\psi}^+ \gamma^0 \psi^+ \,,  \notag \\
F^{a}(e^{b}) &=& R^{a}(e^{b}) + \bar{\psi}^+ \gamma^a \psi^- = de^{a}+\epsilon ^{ac}\omega e_{c}+\epsilon ^{ac}\tau \omega
_{c}-\frac{2}{\ell}\epsilon^{ac}\tau e_{c} + \bar{\psi}^+ \gamma^a \psi^- \,,  \notag \\
F(m) &=& R(m) + \frac{1}{2} \bar{\psi}^- \gamma^0 \psi^- + \bar{\psi}^+ \gamma^0 \rho = dm+\epsilon ^{ac}e_{a}\omega _{c}-\frac{1}{\ell}\epsilon^{ac}e_{a}e_{c} + \frac{1}{2} \bar{\psi}^- \gamma^0 \psi^- + \bar{\psi}^+ \gamma^0 \rho \,,  \notag \\
F(s) &=& R(s) + \frac{1}{\ell} \bar{\psi}^- \gamma^0 \psi^- + \frac{2}{\ell} \bar{\psi}^+ \gamma^0 \rho = ds+\frac{1}{2}\epsilon ^{ac}\omega _{a}\omega _{c} + \frac{1}{\ell} \bar{\psi}^- \gamma^0 \psi^- + \frac{2}{\ell} \bar{\psi}^+ \gamma^0 \rho \,, \notag \\
F(t_1) &=& dt_1 + \frac{1}{\ell} \bar{\psi}^+ \gamma^0 \psi^+ \,, \notag \\
F(t_2) &=& dt_2 - \frac{1}{\ell} \bar{\psi}^- \gamma^0 \psi^- + \frac{2}{\ell} \bar{\psi}^+ \gamma^0 \rho \,, \notag \\
F(u_1) &=& du_1 + \frac{1}{2} \bar{\psi}^+ \gamma^0 \psi^+ \,, \notag \\
F(u_2) &=& du_2 - \frac{1}{2} \bar{\psi}^- \gamma^0 \psi^- + \bar{\psi}^+ \gamma^0 \rho \,, \label{supercurvNRone}
\end{eqnarray}%
with $R(\omega)$, $R^{a}(\omega ^{b})$, $R(\tau )$, $R^{a}(e^{b})$, $R(m)$, and $R(s)$ defined in \eqref{curvaturesNR}, while the fermionic field-strenghts are
\begin{eqnarray}
\nabla \psi^+ &=& d\psi^+ + \frac{1}{2} \omega \gamma_0 \psi^+ + t_1 \gamma_0 \psi^+ \,,  \notag \\
\nabla \psi^- &=& d\psi^- + \frac{1}{2} \omega \gamma_0 \psi^- + \frac{1}{2} \omega^a \gamma_a \psi^+ - t_1 \gamma_0 \psi^- \,,  \notag \\
\nabla \rho &=& d\rho + \frac{1}{2} \omega \gamma_0 \rho + \frac{1}{2} s \gamma_0 \psi^+ + \frac{1}{2} \omega^a \gamma_a \psi^- + t_2 \gamma_0 \psi^+ + t_1 \gamma_0 \rho \,. \label{supercurvNRtwo}
\end{eqnarray}

The supergravity CS action based on the torsional NR superalgebra given by \eqref{torNR} and \eqref{supertorIT} is then obtained by considering the non-vanishing components of the invariant tensor \eqref{torIT} and \eqref{supertorIT} together with the gauge connection one-form \eqref{supertor1f} in the general expression of the CS action \eqref{CS-Action}. By doing so, we get
\begin{eqnarray}
I^{\text{super}}_{\text{NR}} &=&\frac{k}{4\pi}\int \tilde{\alpha}_{0}\left[ \omega _{a}R^{a}(\omega
^{b}) - 2 s R\left( \omega \right) - 4 t_1 dt_2 - \frac{4}{\ell} \bar{\psi}^+ \nabla \rho - \frac{4}{\ell} \bar{\rho} \nabla \psi^+ - \frac{4}{\ell} \bar{\psi}^- \nabla \psi^- \right] \notag \\
&& +2\tilde{\alpha}_{1}\left[
e_{a}R^{a}(\omega ^{b})-mR(\omega )-\tau R(s) - \frac{1}{\ell}e_{a}R^{a}\left( e^{b}\right)+\frac{1}{\ell}mR\left( \tau \right)+\frac{1}{\ell}\tau R\left(m\right) \right. \notag \\
&& +\left. \frac{1}{\ell^2}\tau\epsilon^{ac}e_{a}e_{c} - 2 t_1 du_2 - 2 t_2 du_1 + \frac{4}{\ell} u_1 du_2 -  \bar{\psi}^+ \nabla \rho -  \bar{\rho} \nabla \psi^+ -  \bar{\psi}^- \nabla \psi^- \right] \,. \label{supertorCS}
\end{eqnarray}
The NR CS supergravity action \eqref{supertorCS} contains two independent sectors and can be seen as the NR version of the $\mathcal{N}=2$ relativistic teleparallel CS supergravity action. The first term corresponds to a supersymmetric NR exotic action, while the contribution proportional to $\tilde{\alpha}_{1}$ can be seen as a supersymmetric cosmological extension of the extended Bargmann gravity action \cite{Bergshoeff:2016lwr}. In the vanishing cosmological constant limit $\ell\rightarrow\infty$ we recover the most general extended Bargmann supergravity action \cite{Concha:2020eam}. In particular, the term proportional to $\tilde{\alpha}_{0}$ is no more supersymmetric in the flat limit. Neglecting the fermionic contributions and the additional bosonic gauge fields $\{t_1,t_2,u_1,u_2\}$, \eqref{supertorCS} precisely boils down to the purely NR bosonic action \eqref{torCS} we have previously presented. 

Let us stress that the non-degeneracy of the invariant tensor is preserved as long as $\tilde{\alpha}_0\neq-\frac{\ell}{2}\tilde{\alpha_1}$ and $\tilde{\alpha}_1\neq 0$, which implies that the field equations derived from \eqref{supertorCS} give the vanishing of the curvatures \eqref{supercurvNRone} and \eqref{supercurvNRtwo}. Here, analogously to the purely bosonic case, the cosmological constant can be seen as a source for the spatial super-torsion $\hat{T}^{a}\left(e^{b}\right)=de^{a}+\epsilon ^{ac}\omega e_{c}+\epsilon ^{ac}\tau \omega_{c} + \bar{\psi}^+ \gamma^a \psi^-$ and for the curvature $\hat{T}\left(m\right)=dm+\epsilon ^{ac}e_{a}\omega _{c} + \frac{1}{2}\bar{\psi}^- \gamma^0 \psi^- + \bar{\psi}^+ \gamma^0 \rho$. In particular, on-shell we have
\begin{eqnarray}
\hat{T}^{a}\left(e^{b}\right)&=&\frac{2}{\ell}\epsilon^{ac}\tau e_{c}\,,\notag \\
\hat{T}\left(m\right)&=&\frac{1}{\ell}\epsilon^{ac}e_{a}e_{c}\,.
\end{eqnarray} 
Hence, the torsional NR supergravity theory is then characterized, on-shell, by vanishing supercurvatures $F\left(\omega\right)=0=F^{a}\left(\omega^{b}\right)$ and a non-vanishing spatial super-torsion $\hat{T}^{a}\left(e^{b}\right)\neq0$. In the vanishing cosmological constant limit $\ell\rightarrow\infty$, the field equations reduces to those of the extended Bargmann ones in which the super-torsion vanishes. Let us note that the present theory is quite different from the extended Newton-Hooke supergravity at the dynamic level. Indeed, the Newton-Hooke supergravity theory is characterized by a vanishing super-torsion and the cosmological constant appears explicitly along the supercurvatures of the $\omega$ and $\omega^{a}$ NR gauge fields \cite{Ozdemir:2019tby, Concha:2020eam}. 

As an ending remark, one can alternatively obtain the torsional NR supergravity action \eqref{supertorCS} from the relativistic $\mathcal{N}=2$ teleparallel CS supergravity action \cite{Caroca:2021njq}: 
\begin{eqnarray}
    I_{\text{TSG}}^{\mathcal{N}=2} &=& \frac{k}{4 \pi} \int_{\mathcal{M}} \bigg\{ \alpha_0 \left( W^A d W_A + \frac{1}{3}\epsilon^{ABC}W_A W_B W_B -2 \mathcal{A} d\mathcal{A} -\frac{4}{\ell}\bar{\Psi}^{i}\nabla\Psi^{i}
    \right) \notag\\
    & & + \alpha_1 \Big ( 2 E_A R^A(W) + \frac{4}{3\ell^{2}} \epsilon^{ABC} E_A E_B E_C - \frac{2}{\ell}T^A E_A -4 \mathcal{C}d\mathcal{A} +\frac{4}{\ell}\mathcal{C}d\mathcal{C} \notag \\
    & & -2\bar{\Psi}^{i}\nabla\Psi^{i} \Big ) \bigg\}\,,\label{NtwoTsG}
\end{eqnarray}
where $i=1,2$ and
\begin{eqnarray}
 R^{A}\left(W\right)&=&dW^{A}+\frac{1}{2}\epsilon^{ABC}W_B W_C \,,\notag \\
 T^{A}&=& dE^{A}+\epsilon^{ABC}W_B E_C \,, \notag \\
 \nabla \Psi^i &=& d \Psi^i + \frac{1}{2} W^A \gamma_A \Psi^i + \epsilon^{ij} \mathcal{A} \Psi^j \,. \label{curvsupertel}
\end{eqnarray}
The NR supergravity action \eqref{supertorCS} appears by expressing the NR gauge fields in terms of the relativistic ones through the semigroup elements as
\begin{eqnarray}
\omega&=&\lambda_0 W^{0}\,, \qquad \qquad \omega^a=\lambda_1 W^a\,, \qquad \qquad  s=\lambda_2 W^{0}\,,  \notag \\
h&=&\lambda_0 E^{0}\,, \qquad \ \, \qquad  e^a=\lambda_1 E^a\,, \qquad \, \ \ \quad m=\lambda_2 E^{0}\,,  \notag \\
\psi^{+}&=&\lambda_0 \Psi^{+}\,, \qquad \quad \ \, \ \psi^{-}=\lambda_1 \Psi^{-}\,, \qquad \qquad \rho=\lambda_2 \Psi^{+}\,,  \notag \\
t_1&=&\lambda_0 \mathcal{A} \,, \qquad \qquad \ \ \ t_2=\lambda_2 \mathcal{A}\,, \notag \\
u_1&=&\lambda_0 \mathcal{C} \,, \qquad \qquad \ \ \ u_2=\lambda_2 \mathcal{C}\,, \label{exp2}
\end{eqnarray}
plus the expansion of the relativistic parameters as in \eqref{exp1}. Here, we have considered the decomposition of the $A$ index as in \eqref{decom} along with the following redefinition of the spinor one-form fields:
\begin{equation}
    \Psi^{\pm}=\frac{1}{\sqrt{2}}\left(\Psi^{1}_{\alpha}\pm\left(\gamma^{0}\right)_{\alpha\beta}\Psi^{2}_{\beta}\right)\,.
\end{equation}


\section{Conclusions}\label{sec5}

In this work we have presented the NR counterpart of the teleparallel CS (super)gravity theory in three spacetime dimensions. We first focused on the purely bosonic case, studying the NR limit of the relativistic teleparallel CS gravity \cite{Caroca:2021njq}, based on the teleparallel algebra \eqref{telea}. A torsional NR algebra has been obtained by considering a particular $U\left(1\right)$-enlargement of \eqref{telea}. Such enlargement, with involves additional Abelian generators, is required firstly to ensure a non-degenerate invariant tensor in the limit process to the NR theory and secondly to have a well-defined flat limit leading to the extended Bargmann symmetry. The NR action obtained is characterized, on-shell, by non-vanishing spatial torsion, which is sourced by the cosmological constant. In the vanishing cosmological constant limit $\ell \rightarrow \infty$ such spatial torsion vanishes and the NR CS gravity model boils down to the extended Bargmann gravity theory \cite{Bergshoeff:2016lwr}. The latter was also obtained as the vanishing cosmological constant limit of the CS gravity theory invariant under the Newton-Hooke symmetry. However, as we have shown, the inclusion of the cosmological constant through the torsional NR algebra is different from the one constructed using the Newton-Hooke symmetry. Hence, we have also provided another NR CS gravity model which can be seen as an alternative cosmological extension of the extended Bargmann gravity theory.

At the supersymmetric level, we have considered the three-dimensional relativistic $\mathcal{N}=2$ teleparallel CS supergravity theory \cite{Caroca:2021njq} and developed its NR counterpart by performing the S-expansion procedure \cite{Izaurieta:2006zz}. Indeed, the S-expansion method can be seen as a generalization of the Inönü-Wigner contraction when the semigroup considered belongs to the $S^{(N)}_E$ family. In particular, the semigroup $S^{\left(2\right)}_{E}$ allows to obtain not only a well-defined NR teleparallel (super)algebra but also its invariant tensor.
Analogously to the non-vanishing super-torsion appearing in the relativistic three-dimensional $\mathcal{N}=2$ teleparallel CS supergravity, the torsional NR CS supergravity theory presented here is characterized, on-shell, by a non-vanishing spatial super-torsion sourced by the cosmological constant. In the vanishing cosmological constant limit $\ell\rightarrow\infty$ of the aforementioned NR model, the supersymmetric extension of the extended Bargmann gravity theory \cite{Bergshoeff:2016lwr, deAzcarraga:2019mdn,Concha:2020eam} is recovered. 

To obtain the respective NR version of a relativistic superalgebra is a challenging task and, to our knowledge, has only been approached in three spacetime dimensions \cite{Andringa:2013mma,Bergshoeff:2015ija,Bergshoeff:2016lwr,Ozdemir:2019orp,Concha:2019mxx,Concha:2020eam,deAzcarraga:2019mdn,Ozdemir:2019tby,Gomis:2019nih,Concha:2020tqx,Concha:2021jos,Grassie:2021zgc}. The S-expansion method not only offers us a straightforward mechanism to derive its corresponding NR counterpart but also provides us with the non-vanishing components of the NR invariant tensor. Such procedure could be useful to go beyond three spacetime dimensions and to derive NR CS (super)gravity model in higher odd spacetime dimensions. In even dimensions, a completely different approach should be considered. In this direction, a geometrical formulation of supergravity à la MacDowell-Mansouri \cite{MacDowell:1977jt,Andrianopoli:2014aqa,Andrianopoli:2021rdk} could be used.

Further generalizations of the teleparallel CS gravity theory has been obtained in \cite{Adami:2020xkm} by considering a non-vanishing torsion in the Maxwell CS gravity theory \cite{Cangemi:1992ri,Duval:2008tr,Salgado:2014jka,Hoseinzadeh:2014bla,Caroca:2017izc,Concha:2018zeb,Concha:2018jxx,Concha:2019icz,Chernyavsky:2020fqs}. The study of black hole solutions, and their thermodynamics, of the Maxwellian teleparallel gravity could bring valuable information about the physical implications of a non-vanishing torsion in Maxwell gravity theory [work in progress]. On the other hand, a Maxwellian version of the teleparallel supergravity remains unexplored and could be usefull not only to elucidate the role of torsion in Maxwell supergravity but also to approach a Maxwellian generalization of the present torsional NR supergravity. It would be then interesting to analyze the differences with our torsional NR (super)gravity theory. In particular, one could expect to find a deformation of the Maxwellian extended Bargmann (super)algebra \cite{Aviles:2018jzw,Concha:2019mxx}.


\section*{Acknowledgments}

This work was funded by the National Agency for Research and Development ANID - PAI grant No. 77190078 (P.C.) and FONDECYT grants No. 1211077 (P.C.). This work was supported by the Research project Code DIREG$\_$09/2020 (P.C.) of the Universidad Católica de la Santisima Concepción, Chile. P.C. would like to thank to the Dirección de Investigación and Vice-rectoría de Investigación of the Universidad Católica de la Santísima Concepción, Chile, for their constant support. L.R. would like to thank the Department of Applied Science and Technology of the Polytechnic of Turin, and in particular Laura Andrianopoli and Francesco Raffa, for financial support.


\bibliographystyle{fullsort.bst}
 
\bibliography{NR_supergravity_with_torsion}

\end{document}